\documentclass[3p,times,twocolumn,preprint]{elsarticle}
\usepackage{graphicx}
\usepackage{lineno}

\usepackage[separate-uncertainty=true]{siunitx}
\usepackage{subcaption}
\usepackage{orcidlink}

\usepackage{hyperref}

\usepackage{cleveref}
\crefname{equation}{Eq.}{Eqs.}
\crefname{figure}{Fig.}{Figs.}
\crefname{section}{Sec.}{Secs.}
\Crefname{equation}{Equation}{Equations}
\Crefname{figure}{Figure}{Figures}
\Crefname{section}{Section}{Sections}

\newcommand{\moss}{MOSS}
\newcommand{\most}{MOST} 
\newcommand{\babymoss}{babyMOSS}
\newcommand{\babymost}{babyMOST}

\newcommand{\vs}{\ensuremath{V_\mathrm{s}}}

\DeclareSIUnit\fhrunit{hits/pixel/s}
\DeclareSIUnit\nielunit{1~MeV~n_{eq}~cm^{-2}}
\DeclareSIUnit\electrons{e^{-}}

\begin{document}

\begin{frontmatter}

\title{Testing and Characterization of Wafer-Scale MAPS Prototypes for the ALICE ITS3 Upgrade}

\author[inst,inst2]{Nicolas Tiltmann\,\orcidlink{0000-0001-8361-3467}\corref{auth}, \textit{on behalf of the ALICE Collaboration}}

\cortext[auth]{Corresponding author. e-mail: nicolas.tiltmann@cern.ch}

\address[inst]{European Organisation for Nuclear Research (CERN), Geneva, Switzerland}
\address[inst2]{Universit\"at M\"unster, M\"unster, Germany}

\begin{abstract}
The ALICE experiment will upgrade the innermost three layers of its vertexing detector, the Inner Tracking System (ITS), during the next LHC Long Shutdown (LS3) with a novel, bent, ultra-light MAPS-based tracker. Six wafer-scale sensor chips will be bent into three cylinders, held in place only by carbon foam, leaving no material except for the silicon die in most of the ALICE central barrel acceptance. Two prototype ASICs, approximately $25.9\,\mathrm{cm}$ in length, called MOSS (MOnolithic Stitched Sensor) and MOST (MOnolithic Stitched sensor with Timing), have been produced. These two chips follow complementary approaches to evaluate the use of stitched CMOS sensors for the first time in an HEP experiment.

This article gives an overview of powering tests, functional studies, pixel matrix characterization, and in-beam tests of both test structures. The overall yield of MOSS is measured to be approximately $\SI{76}{\percent}$ per region (1/80th of a chip). This number takes into account powering, as well as functional aspects such as digital and analog pulsing. Two major failure modes have been identified and understood: short in the power grid of the chip and readout issues, that can be clearly attributed to the readout architecture design. Disregarding these issues, the overall yield increases to about $\SI{98}{\percent}$ per region. Furthermore, it is shown that MOSS can operate with $>\SI{99}{\percent}$ efficiency and $<\SI{e-1}{\fhrunit}$ fake-hit rate up to $\SI{4}{\kilo\gray}$ TID and $\SI{4e12}{\nielunit}$ NIEL.


The MOST prototype successfully demonstrated use of power gating, which allows for disconnected parts of the pixel matrix from the power grid in case of shorts. MOSS and MOST successfully proved that designing stitched wafer-scale sensors is feasible and deliver valuable input for the design of the final ITS3 ASIC.

\noindent
Keywords: Tracking, Silicon, MAPS, Stitching, Radiation, LHC    

\end{abstract}
\end{frontmatter}

\section{Introduction }
\label{sec:intro}

\begin{figure}[!h]
  \vspace{.2cm}
  \centering
  \includegraphics[width=.8\linewidth]{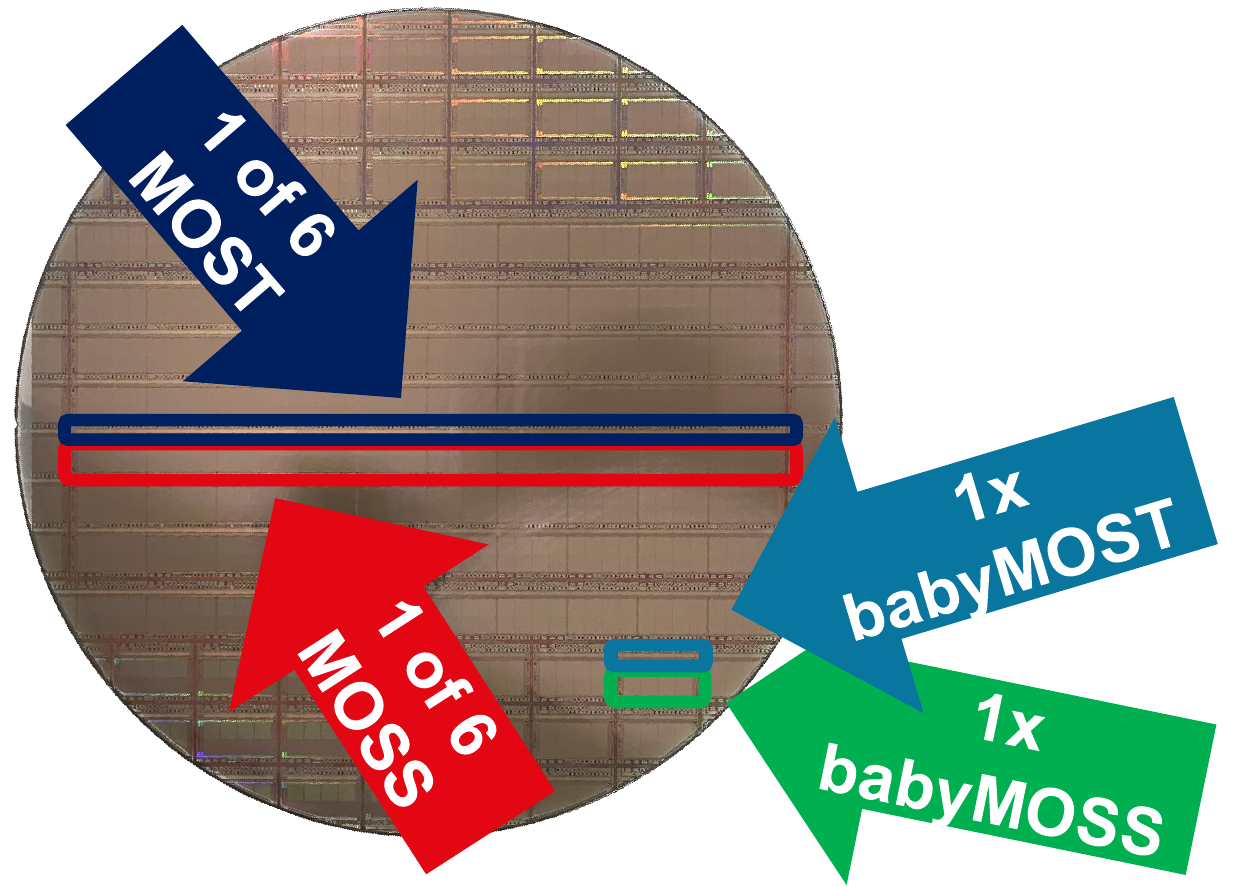}
  \caption{Wafer of the ER1 submission. The smaller boxes show instances of the \babymoss{} and \babymost{} made up of a single RSU and both end caps. \label{fig:waferpicture} }
\end{figure}

In the context of the ALICE ITS3 upgrade project \cite{its3tdr}, a second submission (Engineering Run 2, ER2) has been designed in a \SI{65}{\nano\meter} process. The key feature of this submission are two large scale stitched sensor prototypes. The \textit{MOnolithic Stitched Sensor} (MOSS) \cite{designerpaper} and \textit{Monolithic Stitched Sensor with Timing} (MOST) measure \SI{259}{\milli\meter} in length instrumenting a large fraction of the \SI{300}{\milli\meter} wafer in a single chip. A picture of the actual wafer is presented in \cref{fig:waferpicture}. \moss{} and \most{} follow complementary design approaches in various aspects, which are highlighted in the following.

The \moss{} prototype measures \SI{14}{\milli\meter} in width, while the \most{} prototype is \SI{2.5}{\milli\meter} wide. The pixel pitch of \moss{} is \SI{22.5}{\micro\meter} in the upper half and \SI{18}{\micro\meter} in the lower half, while \most{} employs a single pitch of \SI{18.5}{\micro\meter}.

Both chips share the basic concept of layout, where the central part housing the pixel matrix is a Repeated Sensor Unit (RSU). This RSU is placed 10 times along the long axis of the chip, where interconnection is achieved by stitching. Both short ends of this strip are terminated with separate designs. One of these end caps facilitates the readout of the chip, although the \moss{} can optionally also be read out through pads on the long edge of the sensor, which provides a backup in case of faults in the data transmission backbone along the chip. \most{} has been designed with an integration density at the maximum dictated by this CMOS process, whereas the \moss{} is intentionally designed less dense to possibly assess the impact of integration density in case of insufficient yield.

The powering approach is different in both prototypes. The \most{} pixel matrix is powered by two separate analog power domains and a digital power domain, that are distributed globally across the full chip. As for large chips the probability of having at least one short is significantly higher potentially lowering the overall yield to unacceptable levels, the \most{} implements granular power switches as a mitigation strategy. These switches allow to detach groups of $256$ pixels (analog) or $352$ pixels (digital) from the global power grid allowing for normal operation in the remaining chip. However, issues in the power grid or the switches itself can not be mitigated in this way. To address potential switch-related issues, the \moss{} power grid is organized with increased granularity, with each half-RSU featuring three power domains electrically isolated from the rest of the chip.
\begin{figure}[!h]
    \centering
    \begin{subfigure}[t]{0.22\textwidth}
        \centering
        \includegraphics[width=\textwidth]{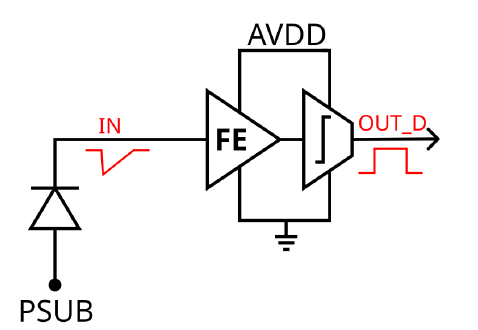}
        \caption{\moss{} biasing scheme.}
        \label{fig:bias_moss}
    \end{subfigure}%
    ~
    \begin{subfigure}[t]{0.22\textwidth}
        \centering
        \includegraphics[width=\textwidth]{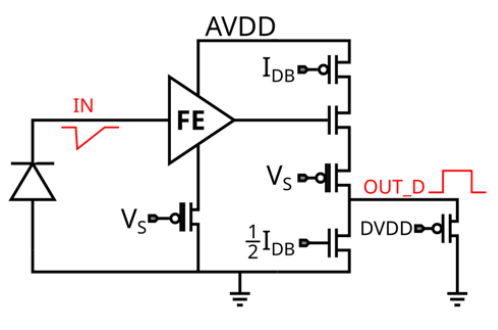}
        \caption{\most{} biasing scheme.}
        \label{fig:bias_most}
    \end{subfigure}
    \caption{Biasing schemes of both large-scale ER1 chips.}
\end{figure}

The diode biasing of \moss{} is realized in a standard way, where a negative voltage can be applied between the substrate and the front-end ground. This is illustrated in \cref{fig:bias_moss}. An alternative approach was chosen for \most{}, where the lower rail of the front-end is shifted up by a voltage $V_\mathrm{s}$ with respect to the global ground, creating a potential difference across the diode. In order to maintain the operational margin of the front-end, the analog supply voltage AVDD also needs to be increased above the nominal \SI{1.2}{\volt}. Shifting back to the regular voltage level is performed in the comparator. A schematic is given in \cref{fig:bias_most}.

The readout of \moss{} is strobed and the pixels possess a hit-latch, which is read out via a priority encoder and additional digital logic in the chip periphery. The \most{} is not strobed and purely hit-driven. 
Upon activation of a pixel, its address is streamed onto one of four global column readout lines shared by all pixels in the column. While this approach enables direct Time-of-Arrival (ToA) and Time-over-Threshold (ToT) measurements, it also introduces a potential risk: if two pixels fire within a short time interval, their signals may collide on the shared readout line, resulting in a sequence that cannot be uniquely decoded.

Along with several other small chiplets, the areas outside of the central rectangle of the wafer are filled up with \babymoss{} and \babymost{} chips, which consist of single RSUs of their respective large versions (see \cref{fig:waferpicture}). The design of the RSU and the end caps is identical to the ones used in the large-scale chips.

\section{Powering Yield}
\label{sec:powering}
\begin{figure}[!h]
    \centering
    \begin{subfigure}[t]{0.23\textwidth}
        \centering
        \includegraphics[width=.95\textwidth]{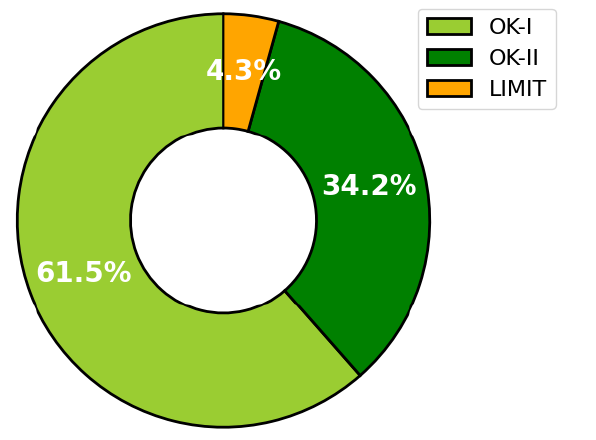}
        \caption{\moss{} yield based on 1620 half-RSUs. Taken from \cite{mosspaper}.}
        \label{fig:moss_yield}
    \end{subfigure}%
    ~ 
    \begin{subfigure}[t]{0.23\textwidth}
        \centering
        \includegraphics[width=.92\textwidth]{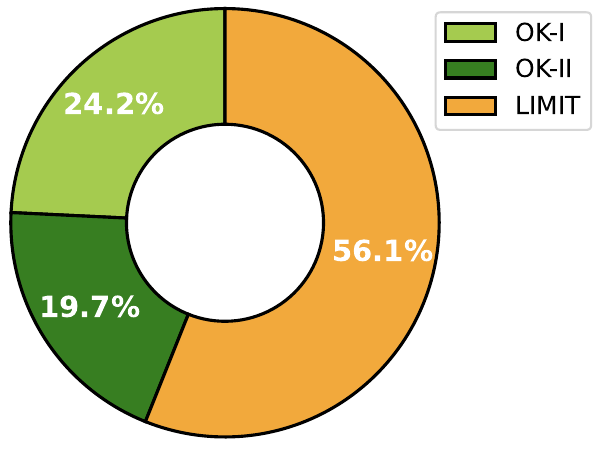}
        \caption{\most{} yield based on 66 full chips.}
        \label{fig:most_yield}
    \end{subfigure}
    \caption{Powering yield of both large-scale ER1 test structures.}
\end{figure}

For the final ALICE ITS3 a low density of critical defects is essential. Conventional detector modules allow for picking the best working chips for assembly. In the ITS3 approach, selecting individual chips is not possible and it is necessary to find six wafers for which a sufficient fraction of units in the center of the wafer are operable. To roughly estimate the yield that can be expected for the final chip (produced in the same \SI{65}{\nano\meter} technology) a mass testing campaign was carried out. A total of 120 \moss{} chips underwent a procedure of cross-impedance measurements between all power nets, followed by slow power ramping under supervision of a thermal camera and another impedance measurement. It is found that thermal hotspots appear during power ramping, which can be correlated with an increased supply current in one of the power domains. In some instances, at a given voltage the current sharply drops to a lower level and continues to rise slowly with increasing voltage normally. In these cases a change of impedance before and after the ramping is observed. Furthermore, the majority of thermal hotspots correlate spatially with the power grid, which leads to the overall hypothesis that shorts are existing in the power grid and that there is a certain probability that a given short is removed (burn-through) by supplying a certain voltage. Importantly, these hotspots occur over all of the chip and not only at the stitching boundaries, indicating no problem with the stitching process. \Cref{fig:moss_yield} shows the distribution of tested half-RSUs. Every half-RSU is categorized into: OK-I -- the half-unit can be powered normally, OK-II -- half-unit can be powered normally, but a transient current was observed, and LIMIT -- a given half-unit can not be powered with a current limit of \SI{500}{\milli\ampere}, where the normal operating currents are in the order of a few tens of \si{\milli\ampere}. In total, only approx. \SI{4.3}{\percent} of half-RSU show persistent shorts and are not operable with the given current limit. All other half-RSUs can reliably and repeatedly be powered, even after burn-through \cite{gregorspaper}.

\Cref{fig:most_yield} shows a similar distribution for \most{}, where OK-I chips show a current below \SI{50}{\milli\ampere} and OK-II chips exceed \SI{50}{\milli\ampere}. LIMIT means that a given chip exceeds a current of \SI{100}{\milli\ampere}. A normally working chip draws approximately \SI{10}{\milli\ampere} per power domain. Due to the global power supply, a tested unit of the \most{} is a full chip, which is about \SI{2.3}{} times as large in area as a \moss{} half-RSU. \SI{56.1}{\percent} of \most{} chips can not powered. Naturally, this number is higher due to the increased chip area. Due to the presence of switches between the power grid and the pixel matrix itself, failures occurring already during power ramping with open switches indicate problems in either the power grid or the switches themselves. The former is compatible with the findings of the \moss{} testing campaign as the general design of the power grid is similar.

\section{Functional Yield}
\begin{figure}[!h]
    \centering
    \includegraphics[width=0.95\linewidth]{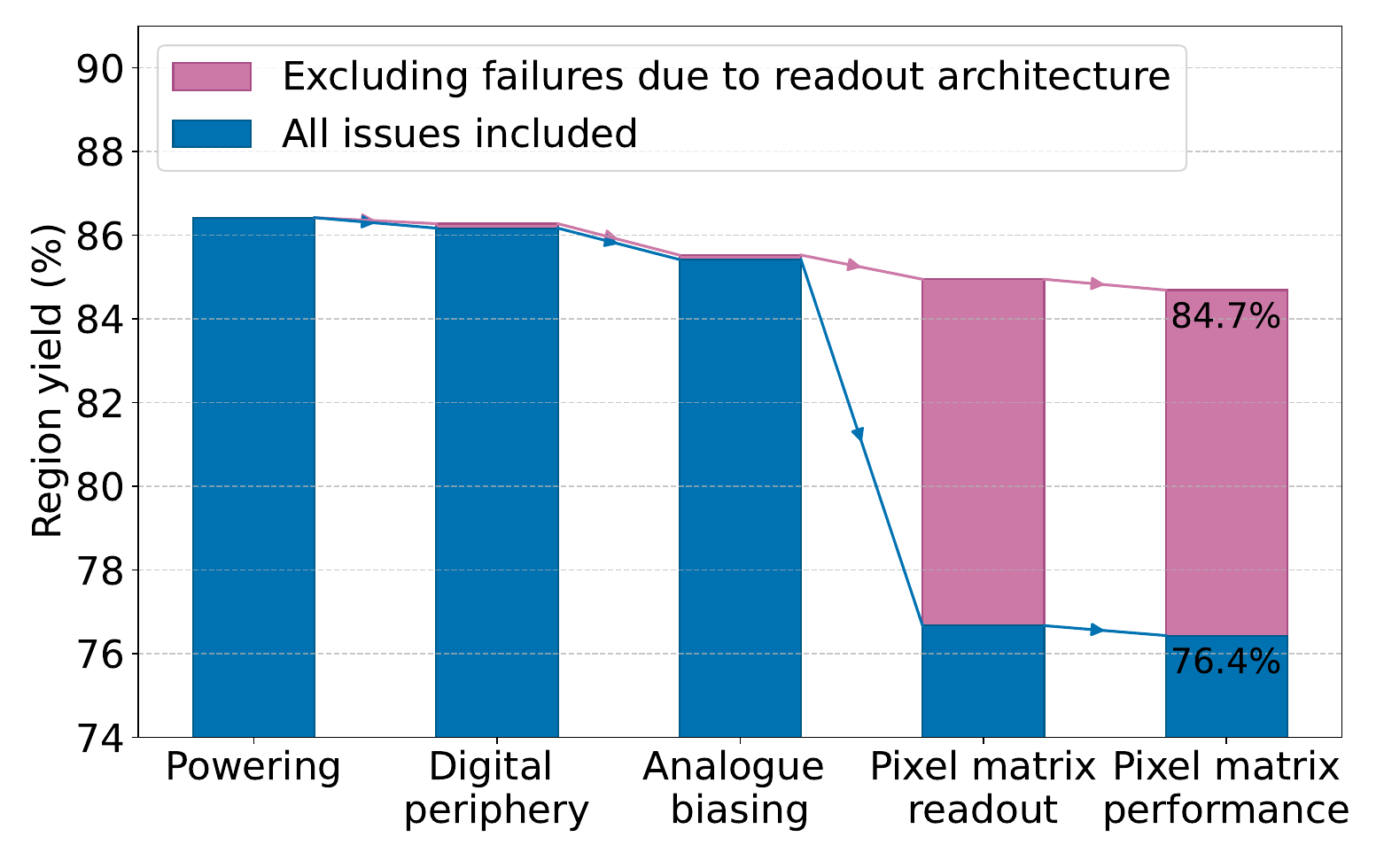}
    \caption{Functional yield of the \moss{} prototype normalized per region. The pink bars illustrate the expected yield when factoring out issues, that are unique to the readout architecture of the prototype and not applicable to future chips. Taken from \cite{mosspaper}.}
    \label{fig:moss_functional_yield}
\end{figure}
The functional yield of \moss{} has been extensively studied in a mass testing campaign of 82 chips from 14 wafers. Testing steps are as follows:
\begin{enumerate}
    \item \textbf{Powering:} Apply power to a unit, apply clock signal, perform reset, set nominal operating point.
    \item \textbf{Digital periphery:} Write to and read back from digital registers.
    \item \textbf{Analog periphery:} Assess linearity of biasing DACs.
    \item \textbf{Pixel matrix readout:} Assert hit latches, then data readout. Afterwards analog pulsing.
    \item \textbf{Pixel matrix performance:} Threshold and noise uniformity evaluation and fake-hit rate measurement.
\end{enumerate}
\Cref{fig:moss_functional_yield} illustrates the achieved yield per region, where one region is a quarter of a half-RSU. From about \SI{86}{\percent} that are passing the powering stage, only about \SI{0.2}{\percent} (\SI{0.1}{\percent}) of all tested units units fail digital (analog) periphery tests. A larger fraction of about \SI{8.1}{\percent} of all tested units are failing the pixel matrix readout test. However, the dominant mechanism of this failure is pixels in which the hit latch can not be deasserted. This problem is specific to the simplistic readout architecture of the \moss{} prototype and such failures are not transferrable to a final chip. The pink bars show the achieved yield where regions with these issues are not considered in calculation. Nevertheless, about \SI{0.6}{\percent} of all regions are failing this step due to $>\SI{1}{\percent}$ dead or noisy pixels, which are in principle transferrable to future chips. Finally approximately \SI{0.3}{\percent} of regions show insufficient threshold or noise uniformity or an excessive fake-hit rate in $>\SI{1}{\percent}$ of pixels, giving a total yield of about \SI{76.4}{\percent}, or \SI{84.7}{\percent} if beforementioned readout architecture issues are removed from the calculation. As discussed in \cref{sec:powering}, the main mechanism of failure for units that can not be powered is understood and will be mitigated. When not considering the powering issue and all issues specific to the readout architecture a total yield of about $\SI{98}{\percent}$ (normalized by region) is achieved \cite{mosspaper}.
The overall functional yield is deemed to be sufficient to construct an ITS3-like detector if one would use sensors similar to \moss{}, disregarding its specific faults and issues \cite{mosspaper}.

Functional yield of \most{} was assessed by analog pulsing of all pixels of a chip. Based on data from 9 chips, it is found that on average \SI{0.009}{\percent} of pixels per RSU are unresponsive \cite{jorysproceedings}.

\section{Energy Calibration}
\begin{figure}[!h]
    \centering
    \includegraphics[width=.4\textwidth]{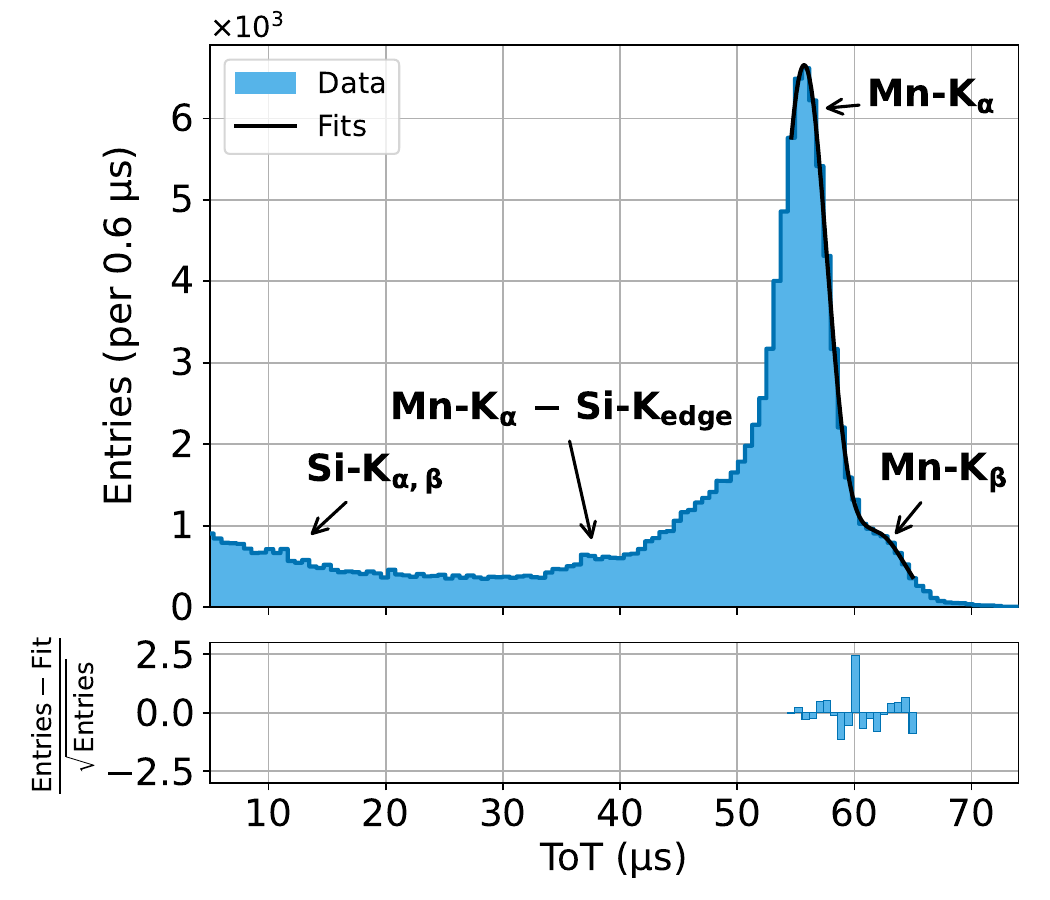}
    \caption{\most{} single pixel cluster spectrum measured with photons from a ${}^{55}\mathrm{Fe}$ source. }
    \label{fig:most_energy}
\end{figure}

Both sensor prototypes are able to measure the energy deposit of incident particles. Here the energy deposit is assessed using a ${}^{55}\mathrm{Fe}$ source emitting photons with energies of $\SI{5.9}{\kilo\electronvolt}$ (Mn-K${}_{\alpha}$) and $\SI{6.5}{\kilo\electronvolt}$ (Mn-K${}_{\beta}$).
\begin{figure}[!h]
    \centering
    \includegraphics[width=.4\textwidth]{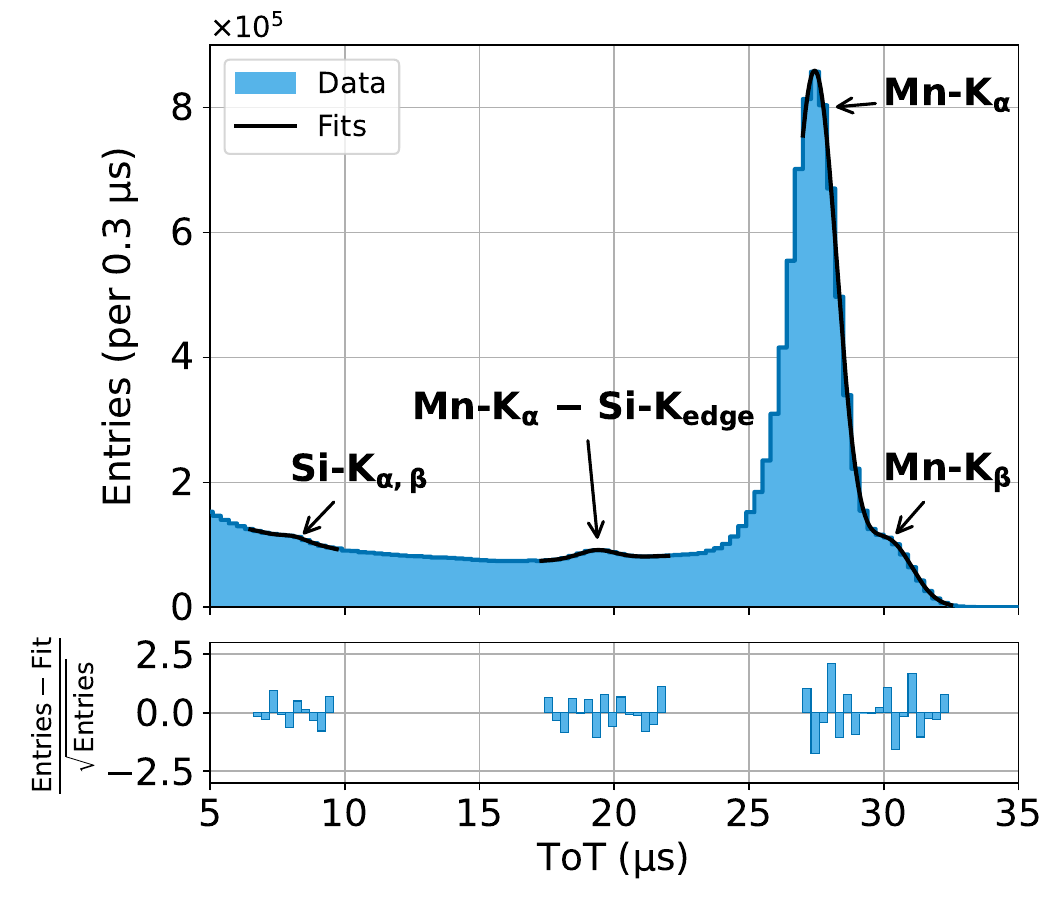}
    \caption{\moss{} single pixel cluster spectrum measured with photons from a ${}^{55}\mathrm{Fe}$ source. Taken from \cite{mosspaper}.}
    \label{fig:moss_energy}
\end{figure}
Pixels of both sensors show linearity between injected charge and ToT. There is a significant pixel-to-pixel variation, which can be compensated for by performing a ToT-vs.-charge calibration pixel by pixel. \Cref{fig:moss_energy,fig:most_energy} show the obtained spectra in units of ToT respectively. In addition to the Mn-K${}_\alpha$ and Mn-K${}_\beta$ peaks, in the case of \moss{} the silicon fluorescence and corresponding escape peak are clearly observable. In the case of \most{} hints of these peaks are identifiable. In both cases, the Mn peaks are fitted with a sum of two Gaussians. The energy resolution is calculated as $\mathrm{FWHM}/\mathrm{mean}=(2\sqrt{2\ln 2}\sigma_{\mathrm{K}_\alpha})/\mu_{\mathrm{K}_\alpha}$, where $\mu_{\mathrm{K}_\alpha}$ and $\sigma_{\mathrm{K}_\alpha}$ are mean and standard deviation of the Mn-K${}_\alpha$ Gaussian fit respectively. This yields an energy resolution of \SI{7.3\pm0.2}{\percent} for \moss{} (\SI{22.5}{\micro\meter} pitch) \cite{mosspaper} and \SI{7.7\pm0.2}{\percent} for \most{}, where only single pixel clusters are considered in both cases.

\section{In-Beam Testing}
\begin{figure}[!h]
    \centering
    \includegraphics[width=.47\textwidth]{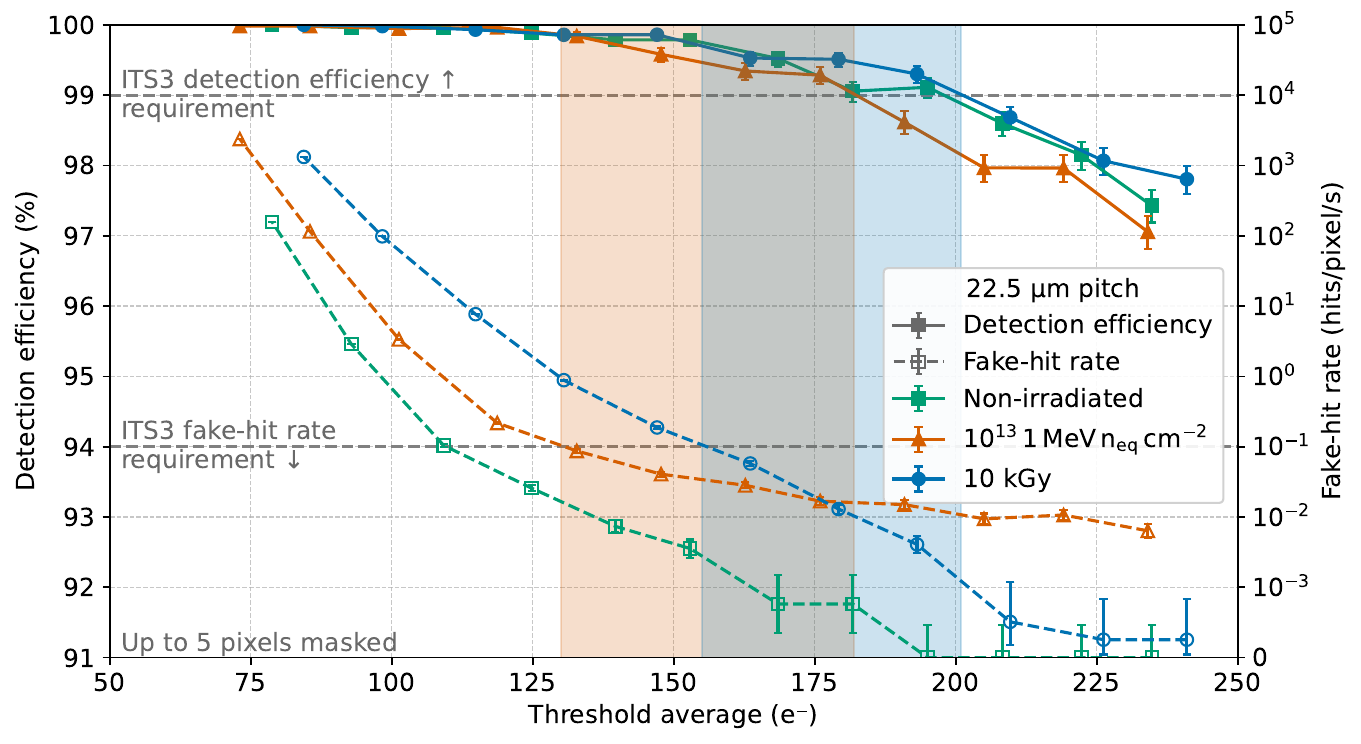}
    \caption{Efficiency and fake-hit rate of \moss{} chips obtained in a beam test at CERN PS. Taken from \cite{mosspaper}, modified.}
    \label{fig:testbeam}
\end{figure}

The \moss{} sensor has been thoroughly studied in beam at CERN PS. \Cref{fig:testbeam} shows the detection efficiency and fake-hit rate as a function of the average sensor threshold for the larger pixel pitch of \SI{22.5}{\micro\meter}. Green curves show a non-irradiated sample. Above $\SI{99}{\percent}$ efficiency and below $\SI{e-1}{\fhrunit}$ are achieved for a range of about \SI{80}{\electrons} of thresholds. Two other samples have been irradiated to an ionizing dose of \SI{10}{\kilo\gray} using x-rays or a non-ionizing dose of \SI{e13}{\nielunit} using nuclear reactor neutrons respectively. As expected, the fake-hit rate increases after exposure to ionizing radiation, which likely is related to a worse performance of the front-end due to altered transistor parameters. The detection efficiency decreases after exposure to non-ionizing radiation. Furthermore, the fake-hit rate increases due to increased leakage current as well as residual radioactivity in the chip. However, an operational margin still remains after ionizing and non-ionizing radiation to ITS3 levels, as depicted by the orange and blue boxes in the background \cite{mosspaper}.

\section{Diode Biasing}
\begin{figure}[!h]
    \centering
    \includegraphics[width=.45\textwidth]{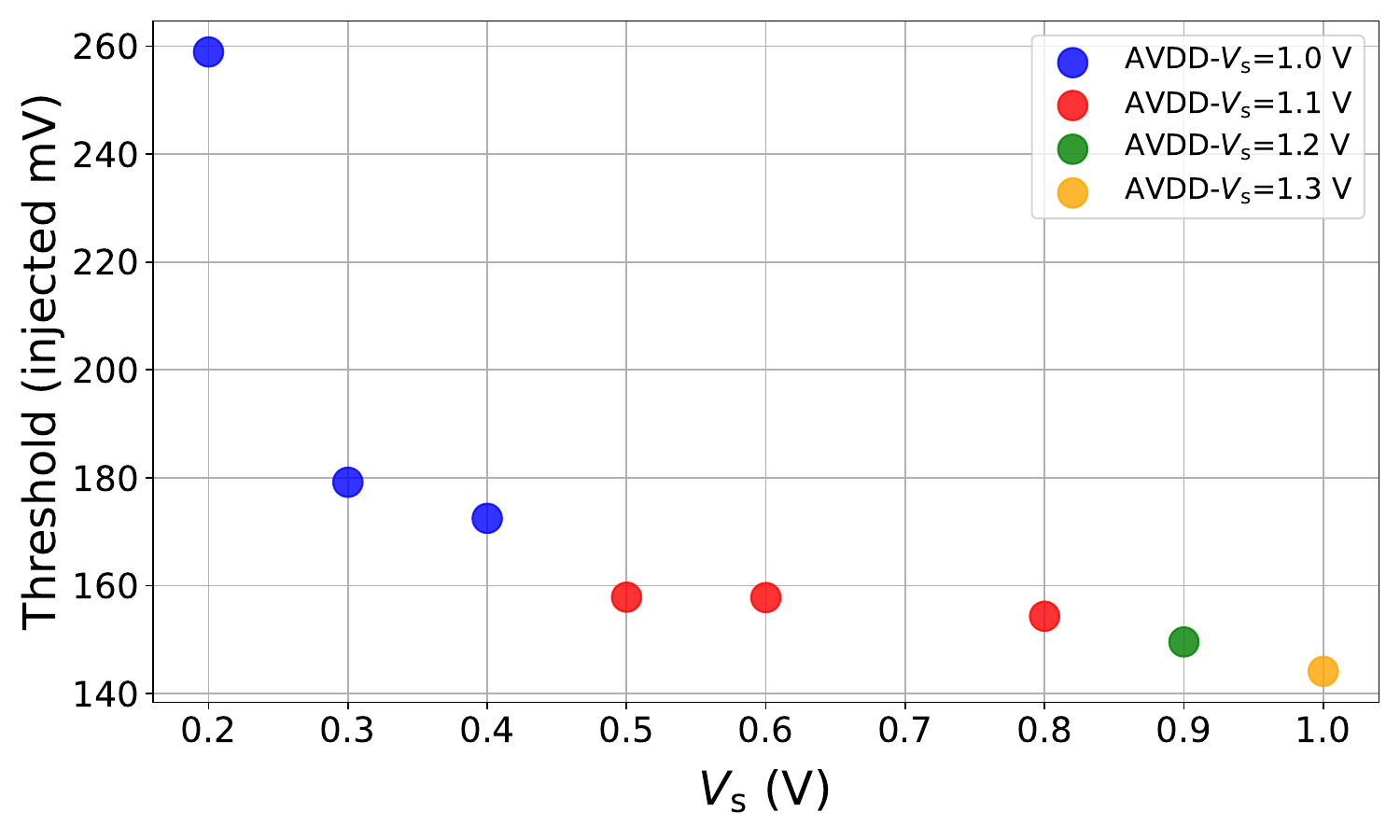}
    \caption{Sensor threshold average in units of injection voltage as function of \vs{} at different AVDD settings for \most{}. Taken from \cite{jorysproceedings}, modified.}
    \label{fig:diodebiasing}
\end{figure}

The alternative biasing scheme of \most{} (c.f. \cref{sec:intro}) has been tested by applying different combinations of AVDD and \vs{} and measuring the sensor threshold. As \vs{} defines the potential difference between the lower rail of the front-end and thus the collection electrode, increasing \vs{} means effectively applying a reverse bias to the sensor diode. AVDD is only increased to remain a sufficient operating margin of the front-end circuitry. The resulting curve is shown in \cref{fig:diodebiasing}, indicating a clear decrease of sensor threshold with increased \vs{}. For future chips this biasing scheme provides an alternative as it removes the need to distribute negative voltages across the chip \cite{jorysproceedings,mariasproceedings}.

\section{Power Gating}
\begin{figure}[!h]
    \centering
    \includegraphics[width=.4\textwidth]{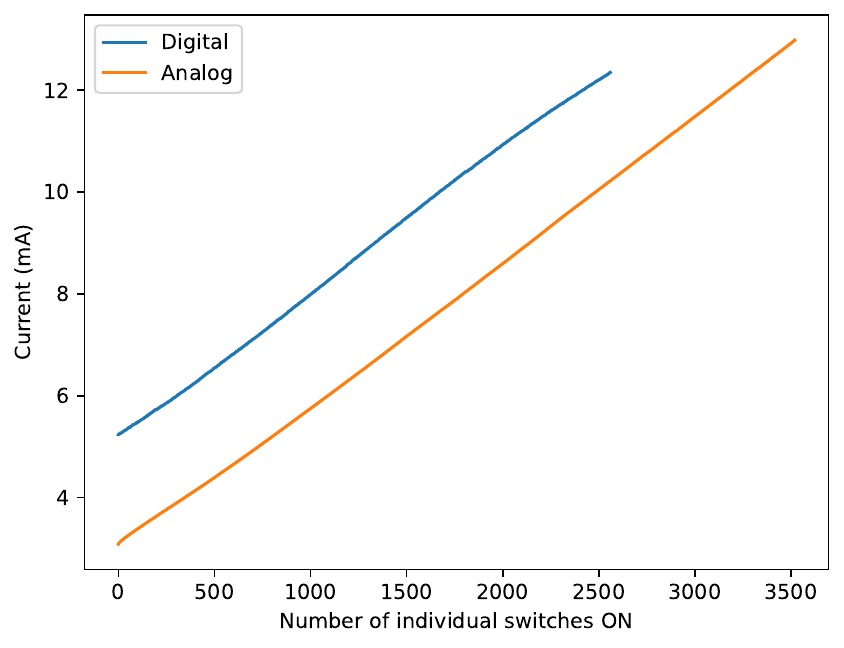}
    \caption{\most{} analog and digital current consumption during subsequent attaching of groups to the power grid via switches.}
    \label{fig:powergating}
\end{figure}

To test the power gating switches, digital and analog supply currents are measured while subsequently closing all individual switches. A smooth uniform increase is observed, as shown in \cref{fig:powergating}. No sudden jumps are occurring, indicating no faulty pixel groups in this chip. Also no plateaus exist, showing that all switches are closing. The effectiveness of a power switch for isolating a fault and operating the rest of the chip could not be proven as in all tested chips, not a single instance of fault has been found, that appeared only after closing the power switches. All observed faults are attributable to shorts in the power grid, for which the power switches do not provide mitigation.

\section{Conclusion}
\moss{} and \most{} have successfully proven that building a wafer-scale stitched sensor is feasible. No problems attributed to the stitching itself have been observed. At the required radiation levels of the ITS3 (\SI{4}{\kilo\gray} and \SI{4e12}{\nielunit}) \moss{} remains operable with $>\SI{99}{\percent}$ efficiency and a fake-hit rate $<\SI{e-1}{\fhrunit}$. The yield achieved is sufficient for constructing the detector, when disregarding issues related to the prototype readout architecture, as well as power grid related faults, which are understood and will likely be mitigated in the next sensor. In this case, a region yield of $\approx\SI{98}{\percent}$ can be achieved. The test features and operational experience of both chips provided valuable feedback for the design of the final ITS3 sensor ASIC.

\section*{Acknowledgements}
This work has been sponsored by the Wolfgang Gentner Programme of the German Federal Ministry of Education and Research (grant no. 13E18CHA).


{}

\end{document}